\documentclass[twocolumn]{aastex62}

\newcommand{\atnf}{PSRCAT}

\usepackage{amsmath, amssymb}

\newcommand{\msun}{M$_{\odot}$}
\newcommand{\be}{\begin{equation}}
\newcommand{\ee}{\end{equation}}

\usepackage{soul}

\shorttitle{}
\shortauthors{M. L. Jones et al.}

\begin{document}

\title{Constraints on Undetected Long-Period Binaries in the Known Pulsar Population}

\correspondingauthor{}
\email{megan.jones@nanograv.org}

\author[0000-0001-6607-3710]{Megan~L.~Jones}
\affil{Center for Gravitation, Cosmology and Astrophysics, Department of Physics, University of Wisconsin-Milwaukee, Milwaukee, WI 53201, USA}

\author[0000-0001-6295-2881]{David~L.~Kaplan}
\affil{Center for Gravitation, Cosmology and Astrophysics, Department of Physics, University of Wisconsin-Milwaukee, Milwaukee, WI 53201, USA}

\author[0000-0001-7697-7422]{Maura A. McLaughlin}
\affiliation{Department of Physics and Astronomy, West Virginia University, P.O. Box 6315, Morgantown, WV 26506, USA}
\affiliation{Center for Gravitational Waves and Cosmology, West Virginia University, Chestnut Ridge Research Building, Morgantown, WV 26505, USA}

\author[0000-0003-1301-966X]{Duncan~R.~Lorimer}
\affil{Department of Physics and Astronomy, West Virginia University, P.O. Box 6315, Morgantown, WV 26506, USA}
\affil{Center for Gravitational Waves and Cosmology, West Virginia University, Chestnut Ridge Research Building, Morgantown, WV 26505, USA
}

\begin{abstract}

Although neutron star-black hole binaries have been identified through mergers detected in gravitational waves, a pulsar-black hole binary has yet to be detected. While short-period binaries are detectable due to a clear signal in the pulsar's timing residuals, effects from a long-period binary could be masked by other timing effects, allowing them to go undetected. In particular, a long-period binary measured over a small subset of its orbital period could {manifest via time derivatives of the spin-frequency incompatible with isolated pulsar properties}. 
We assess the possibility of pulsars having unknown companions in long-period binaries and put constraints on the range of binary properties that may remain undetected in current data, but that may be detectable with further observations. 
We find that 
{for 35\% of {canonical} pulsars with published higher order derivatives, the precision of measurements is not enough to confidently reject binarity} 
(period $\gtrsim2\,$kyr), and that a black-hole binary companion could not be ruled out for {a sample of} pulsars without published constraints if the period is $> 1$\,kyr.
While we find no convincing cases in the literature, we put more stringent limits on orbital period and longitude of periastron for the few pulsars with published higher-order frequency derivatives {($n \geq 3$)}. 
We discuss the detectability of candidates and {find that a sample pulsar in a 100\,yr orbit could be detectable within 5$-$10\,yr}.

\end{abstract}


\section{Introduction}

Pulsar-black hole (BH) binaries have been described as the ``holy grail of astrophysics" \citep{fau11}; the discovery of such a system could reveal a wealth of information on stellar evolution and strong gravity. Although the exact number is highly uncertain, predictions range from a few to roughly 100 such binaries existing in the Galaxy \citep{lip05,lor12,shao18}. While no pulsar-BH binaries are yet known, large-scale pulsar surveys with the SKA are predicted to increase the known pulsar population to 30,000 sources, {3000 of which are expected to be recycled pulsars} \citep{ska}, greatly increasing the likelihood of discovering exotic pulsar binaries such as these.

\cite{sig03} outlines two formation scenarios for a pulsar-BH binary in the Galactic field. (1) The system starts as a binary consisting of two massive stars. The primary, more massive star explodes, forming a BH with the system remaining bound. The secondary, less massive star then explodes as a supernova, leaving behind a canonical pulsar in a wide orbit. A possible example of a progenitor to such a system is VFTS 243, which hosts a BH ($M_{\rm{BH}} > 9$\,\msun) in a $\sim$10\,day orbit with a 25\,\msun O-type star \citep{ste22}. (2) The pulsar would form first in a tighter binary, outlined in more detail by \cite{sip04}. Both progenitor main sequence stars would be just below the mass threshold to form a BH and would need to be close enough for mass transfer. The neutron star progenitor (i.e.,\, the primary star) would lose enough mass to the secondary star such that the secondary reaches the threshold of BH formation with the primary retaining enough mass to become a neutron star. \cite{sip04} assume that the secondary's mass gain shortens its evolutionary time-scale in accordance with the mass gained, while the primary's lifetime is not extended by mass loss. After the primary forms a neutron star, the two stars need to stay close enough for a second round of mass transfer, this time with the secondary losing mass to the newly formed neutron star. After this second period of mass transfer, the system would contain a recycled, millisecond pulsar (MSP) that will remain bound following the subsequent formation of the BH \citep{tau12}. {High-mass companions that have yet to become BHs could be detected via their optical emission \citep{igo19,ant20,ant21}.} 

The emergence of multi-messenger astronomy has led to additional focus on a range of relativistic binaries incorporating neutron stars \citep[e.g.,][]{mar21}. The LIGO/Virgo experiment has identified several gravitational wave sources consistent with the merger of a neutron star with a stellar-mass BH \citep[e.g.,\,][]{abbott20a,abbott20b}, but a local pulsar-BH binary has yet to be detected at any orbital period. \cite{pol21} predict the number of potentially detectable short-period binaries {($P_b \leq 10$\,days)} in the Galaxy based on LIGO/Virgo estimates of event-based and population-based merger rates, finding a range between $\sim$1 and $\sim$13 detectable binaries. Here, in contrast, we investigate the possibility of a long-period binary {($P_b \lesssim 2$\,kyr)} hiding in the known pulsar population. 

The ATNF Pulsar catalog {v1.65} \citep[PSRCAT\footnote{{https://www.atnf.csiro.au/research/pulsar/psrcat/}};][]{psrcat} lists $\sim$3000 rotation-powered pulsars, $\sim$10\% of which are known to be in binary systems. The longest ``closed" orbit (with a data span longer than the binary period) listed is 5.3\,yr \citep[PSR~J1638$-$4725;][]{lor06}; however, longer period binaries have been identified through anomalous timing behavior. PSR~J2032+4127 is in a highly eccentric orbit ($e = 0.978$) with a 40$-$50\,yr period \citep{lyn15b,ho17}. PSR~B1620$-$26 is in a hierarchical triple system, orbiting a white dwarf with a 191\,day period and a Jupiter-like planet with a $\sim60$\,yr period \citep{tho99}. PSR J1024--0719 is in a long-period binary orbit with a low-mass main-sequence star, identified via an unusually low spin period derivative; estimates on the orbital period are from 0.2 -- 20 \,kyr \citep{bas16,kap16}. The discovery of previously undetected long-period binary companions could explain anomalous timing behavior in other systems. Such a system could also help in understanding varied formation channels for neutron star binaries. This work is particularly timely given modern optical surveys and greatly increased pulsar timing programs through instruments like the Canadian \ion{H}{1} Mapping Experiment (CHIME; \citealt{chime}), the upgraded Molonglo Observatory Synthesis Telescope (UTMOST; \citealt{low20}), and the future Deep Synoptic Array (DSA-2000; \citealt{dsa2000}).

The pulsar timing model characterizes time delays and effects in the data, fitting more parameters depending on the complexity of the system \citep{lor12}. If a particular effect is unmodeled, its resulting delays can be absorbed by other fit parameters, with the power reduced by this fitting quantified by the transmission function \citep[e.g., ][]{bla84}. While short-period binaries would be detectable due to a clear periodic signal in the timing residuals, effects from a long-period binary could be masked by other timing delay effects (e.g.,\, pulse spin-down and timing noise), allowing them to go undetected \citep[e.g.,\,][]{kap16, bas16}. A long-period binary measured over comparably short timescales that is unaccounted for in the timing model could instead be fit out through a sum of polynomial terms (i.e., frequency derivatives).

In this paper, we assess the possibility of unknown long-period pulsar binaries in the pulsar catalog. We find no particularly convincing cases, and put constraints on the range of binary properties that may remain undetected in current data, but that may be detectable with further observations. We derive relations to known pulsar parameters to constrain possible hidden systems in \S2. We put limits on orbital period and companion mass using the pulsar magnetic field and published constraints on frequency derivatives while incorporating a transmission function to account for power loss to other parameters in the timing model in \S3. We discuss the detectability of candidates and plans for future candidate confirmation in \S4.

\section{Method}

For an apparently isolated pulsar, the time-dependent phase of the pulse due to spin-down can be expressed as 
\begin{eqnarray}\label{eq:phase}
    \phi(t) &=& f(t-t_0) +\frac{1}{2}\dot{f}(t-t_0)^2 + \ldots \nonumber\\
    &=& \sum_{n=0}^{\infty} \frac{f_{n}}{(n+1)!} ~(t-t_0)^{(n+1)}~,
\end{eqnarray}
where $t$ and $t_0$ are the current and reference times respectively, and $f_{n}$ is the $n$-th timing frequency derivative \citep[e.g.,][]{lor12}. The inclusion of a binary model adds additional terms to the phase, with the binary model amplitude characterized as $\propto (a/c)/P_b$ where $a$ is the semi-major axis of the binary orbit and $P_b$ is the binary period. The amplitude of binary motion will be quite small as binary period increases, making its effect on pulsar timing residuals harder to detect. 

Any timing parameters that are present but unmodeled in the data will have power absorbed by other parameters in the model; depending on the scale {and shape} of the residuals due to a particular effect, residuals due to that parameter may be completed absorbed. This absorption not only removes traces of this effect, but also incorrectly skews other parameters in the timing model to account for this unmodeled effect. In the case of unmodeled binary motion, a sufficiently small binary amplitude may be completely absorbed by other effects, like pulse spin-down and timing noise. Here we discuss the most likely effects that may absorb the unmodeled binary amplitude.

\subsection{Intrinsic Timing Variations}

The frequency derivatives can contain contributions from both deterministic and stochastic variations for an isolated pulsar. For instance, there are expected to be deterministic frequency derivatives intrinsic to the pulsar arising from magnetic dipole radiation (i.e., pulsar spin-down). While $f_1$ due to spin-down can typically be easily measured via timing, this is usually only possible for $f_2$ for very young pulsars due to their comparatively higher $f_2$ values. For other pulsars, we can predict $f_2$ due to {dipole radiation} as
\begin{equation}\label{eq:dipole}
    f_{2;\rm{d}} = n_b\frac{f_{1}^2}{f_{0}}~,
\end{equation}
where $n_b$ is the braking index; here we assume the canonical value $n_b=3$ \citep[e.g.,][]{lor12,esp17}, although actual measurements of young pulsars typically find smaller values {\citep[e.g.,][]{chen06,par19}}. Higher order spin-down terms can be calculated by taking more derivatives of this relation with respect to $f_0$ \citep{par19}.  

Stochastic spin-down variations known as ``timing noise" --- often referred to as ``red noise" because of its spectral content \citep[e.g.,][]{gon21} ---  can be modeled in a variety of ways, including power-law noise processes \citep[e.g.,][]{las15}.  However, it can also be modeled via a polynomial basis function, leading to additional contributions to $f_2$ and other terms. Like dipole spin-down, this effect is typically stronger in younger pulsars. 

There have been numerous analyses characterizing timing noise in the literature. \cite{arz94} defined the timing stability parameter as 
\begin{equation}
    \Delta_T = \log_{10}\left(\frac{|f_2|}{6f_0} T^3 \right)~,
\end{equation}
where the nominal observation length is $T=10^8$\,s, giving the commonly-reported quantity $\Delta_8$ characterizing the strength of the timing noise as essentially the logarithm of the anticipated rms timing noise as parameterized by $f_2$. 

\cite{sha10} asserted the need for more robust diagnostics to characterize timing noise. After a second order fit for $f_0$ and $f_1$, the authors calculated the post-fit rms $\sigma_{\rm TN}$, which is assumed to be due to timing noise, through maximum likelihood analysis as 
\be\label{eq-corshan}
    \sigma_{\rm{TN}} (\mu s)  = \exp(C) f_0^{\alpha} \left|\frac{f_1}{10^{-15}~s^{-2}}\right|^{\beta} T_{\rm{yr}}^{\gamma}~,
\ee
where $T_{\rm{yr}}$ is the observing time in years, and the best-fit parameters are $C = 2.0 \pm 0.4$, $\alpha = -0.9 \pm 0.2$, $\beta = 1.00 \pm 0.05$, and $\gamma = 1.9 \pm 0.2$. 

\cite{par19} performed a a linear least-squares regression analysis assuming a 10\,yr timing baseline to fit the timing noise metric
\be
    \sigma_{\rm{TN}} (\mu s) = f_0 |f_1|^b ~,
\ee
where $b = -0.9 \pm 0.2$ for canonical pulsars, which agrees with the relation found by \cite{sha10} by rotational symmetry \citep{jan18}. 

For convenience when assessing the full pulsar population, we combine Eqn.~\ref{eq-corshan} with the standard expression for the dipolar magnetic field \citep{lor12}:
\be
B = 3.2\times10^{19}\sqrt{\frac{| f_1 |}{f_0^3} }~,
\ee
to express the anticipated timing noise in terms of the measured $f_0$ and $B$ 
\begin{eqnarray}\label{eq:tn}
    \sigma_{\rm{TN}}   (\mu s)   = 7.22 ~f_0^{2.1}~\left(\frac{B}{10^{12}~\rm{G}}\right)^2~T_{\rm{yr}}^{1.9}~.
\end{eqnarray}

{Beyond the stochastic but continuous timing noise discussed above, pulsars can exhibit discontinuous changes in their spin evolution known as ``glitches" \citep{radhakrishnan1969,reichley1969,fuentes2017}, and unseen glitches may contribute to incorrect estimates of timing noise or other variations in spin-rate \citep[e.g.,][]{antonelli2023,par20,lower2021}, especially in younger/more energetic pulsars.  For instance, the finite sampling of timing programs could miss glitches, and these could to incorrect estimates of the braking index \citep{esp17}.  Changes in the pulse shape \citep[e.g.,][]{singha21} can also mimic stochastic timing variations.  For the most part, the empirical timing noise relations given above likely include some small unmodeled glitches \citep{par19,par20}; larger glitches can be identified separately with sufficient timing cadence.}

\begin{figure}
    \epsscale{1.2}
    \plotone{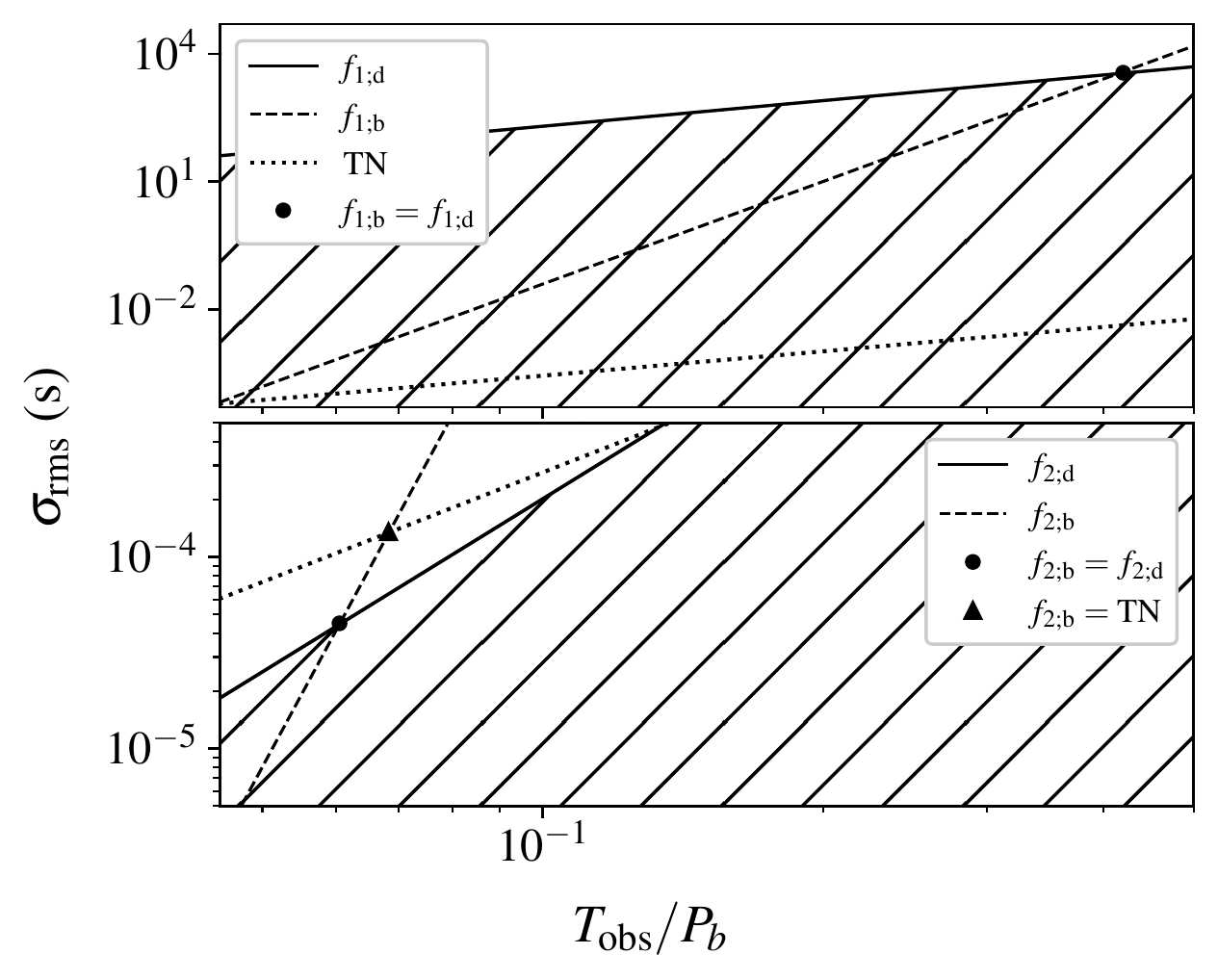}
    \caption{Induced rms residuals over observation span $T$ for various effects. We assume $f_0=$ 10\,Hz, $f_1 = 10^{-15}$\,Hz/s, $B = 3.2 \times 10^{10}$\,G, $f_2$ calculated using Eqn.~\ref{eq:dipole}, $M_{\rm{BH}}$ = 10\,M$_{\odot}$, and $P_b =  100$\,years. The top panel shows the comparison for $f_1$ while the bottom panel compares different contributions to $f_2$; a comparison to timing noise can be seen for both. The hatched regions show where residuals due to an unmodeled circular, edge-on binary companion will be overwhelmed by spin-down. The timing residuals from {$f_1$ due to binary motion} $f_{1;b}$ surpass those from {$f_1$ due to dipole radiation} $f_{1;d}$ around $\sim0.4P_b$, whereas those from $f_{2;b}$ surpass those from $f_{2;d}$ after $\sim0.06P_b$, making the search for a binary using $f_{2}$ more feasible than using $f_1$. } 
    \label{fig:sigmas}
\end{figure}

\subsection{Extrinsic Timing Variations}
{Torque variations, whether inside the neutron star or in the magnetosphere \citep{antonelli2023} will contribute to higher spin derivatives of neutron stars.  There also can be external causes for timing variations.  For instance, changes in the interstellar dispersion \citep{lam15,jones17} can contribute to timing noises.  However, here we focus on the presence of an unmodeled binary companion.}

When not explicitly characterized by the timing model, unmodeled binary motion will introduce pulse residuals represented as sinusoids (for circular orbits); however, if $T\ll P_b$ then the residuals can instead be modeled as pulse frequency derivatives \citep{jos97}. These derivatives can be expressed via Doppler shifts (and higher order terms)
\begin{equation}\label{eq:fn}
    f_{n;b} = \left(\frac{2\pi G \tilde{M}}{P_b c^3}\right)^{1/3} \left(\frac{f_0}{n!}\right) \left(\frac{-2\pi }{P_b}\right)^{n} 
    \sin(\omega + \frac{n \pi}{2})
    \sin{i}~, 
\end{equation}
where $\tilde{M}$ = $M_{\rm{BH}}^3/(M_{\rm{BH}} + M_P)^2$ with $M_{\rm{BH}}$ as the BH mass, $M_P$ is the pulsar mass, $\omega$ is the longitude of periastron, 
and $i$ is the binary inclination \citep[e.g.,][]{bha08}. This expression for $f_{n;b}$ can be used with Eqn.~\ref{eq:phase} to calculate the induced rms timing residuals $\sigma_{\rm rms}\approx\langle \Delta\phi(t)^2 \rangle^{1/2}/f_0$. Note that Eqn.~\ref{eq:fn} assumes a circular binary for simplicity, whether or not this is a good physical assumption; given the formation mechanisms discussed earlier, we would likely not expect a circular binary \citep[e.g.,][]{dec15}. An eccentric system can be straightforwardly approximated by adding an additional eccentricity factor $\epsilon = (1 + e \cos{(\lambda)}/(1-e^2)$, where $\lambda$ is the longitude of the companion from pericenter \citep{jos97}. 

For all pulsars, we specifically look for {canonical} sources that are not in a known binary and are not associated with a globular cluster, as cluster dynamics could introduce their own frequency derivative components \citep[e.g.,][]{phi93}. While present, contributions to $f_2$ due to Shklovskii acceleration, Galactic motion, and differential acceleration will not sufficiently contribute to the frequency derivatives for canonical pulsars considered here, but are likely relevant effects for MSPs \citep{shk70,gui16,liu18,pat21}.

\subsection{Intrinsic Versus Extrinsic Frequency Derivatives}

\begin{figure}
    \epsscale{1.2}
    \plotone{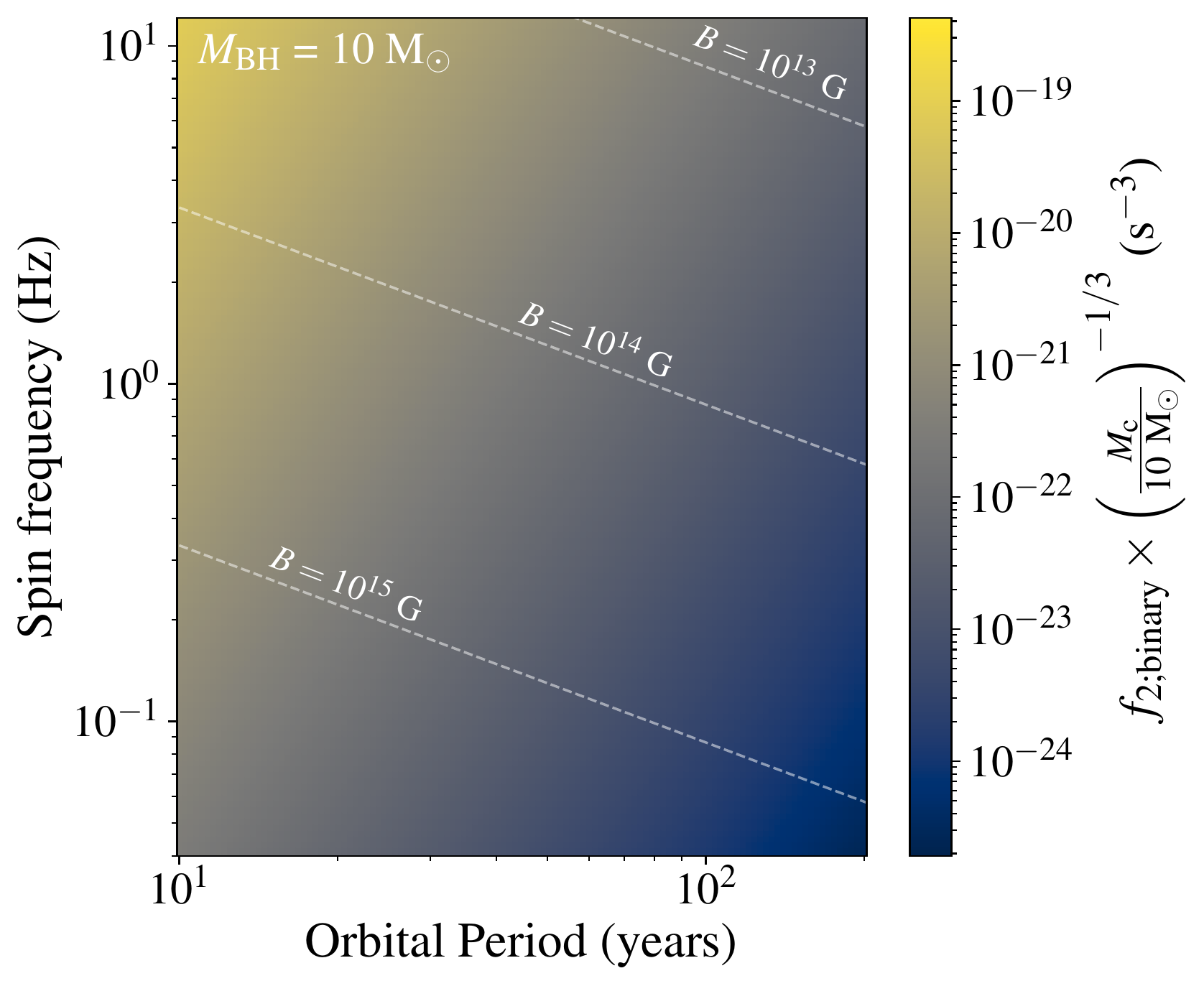} 
    \caption{The $f_2$ induced for a circular binary for a range of spin and orbital periods. The black hole companion mass is assumed to be 10\,$M_{\odot}$, and we assume the binary orbit is edge-on ($i=90^{\circ}$). The dashed lines represent the surface magnetic field $B$ corresponding to a specific $f_2$ due solely to dipole radiation ($n_b=3$). To use this for a particular pulsar with a single value of $f_0$ and $B$, any orbital period to the left of where that $f_0$ value crosses the appropriate $B$ contour may be probed in binary searches. Following Eqn.~\ref{eq:upperlimit}, this suggests an upper limit on orbital period where longer periods will be hidden by magnetic field effects. {The magnetic fields necessary to feasibly attribute unmodeled binary motion to dipole radiation are generally higher than would be expected for most canonical pulsars.}
    }
    \label{fig:map}
\end{figure}

When looking for unmodeled binary motion (Eqn.~\ref{eq:fn}), we must compare the expected frequency derivatives from such binaries to those expected from intrinsic sources (spin-down and timing noise, to start). A comparison of the induced timing residuals over time for dipole radiation, intrinsic timing noise, and an unmodeled binary can be seen in Fig.~\ref{fig:sigmas}. For the first frequency derivative, $f_1$ due to a binary would be overwhelmed by spin-down effects \citep[][and Fig.~\ref{fig:sigmas}]{lamb76}. We therefore focus on the second derivative $f_2$ as it is the first term that is potentially larger than the dipole radiation contribution. For a binary to be detectable, the orbital period must be short enough that the induced timing residuals stand out with respect to other sources of timing offsets, but long enough that the binary motion is not obvious from the residuals (i.e., $T\ll P_b$). Predicted $f_2$ values for various spin frequencies and orbital periods can be seen in Fig.~\ref{fig:map} along with the necessary pulsar magnetic field to account for the induced $f_2$ due to a binary orbit. 

\section{Limits for known pulsars}

Given the constraints on binary detectability via frequency derivatives as well as intrinsic timing effects, we determine an orbital parameter phase space in which we are sensitive for the known pulsar population. We will determine these limits on detectability through several methods in the next sections depending on the highest order frequency derivative published for a given pulsar that meets these criteria.

\subsection{Sample Selection}\label{sec-sampleselec}

\subsubsection{Pulsars without higher order constraints}

First, we use the calculated magnetic field as the upper limit for pulsars without published higher order frequency derivatives to predict $f_2$ due to dipole radiation in the absence of a measured constraint; induced $f_2$ from binaries with periods above this limit will be overwhelmed by spin-down effects. Combining Eqns.~\ref{eq:dipole} and \ref{eq:fn} yields an upper limit constraint on orbital period
\begin{eqnarray}\label{eq:upperlimit}
    P_b^{7/3} & ~\lesssim ~&  \frac{1}{n_b f_0^4}  \left(\frac{\tilde{M}}{10~ M_{\odot}}\right)^{1/3} \left(\frac{B}{10^{15}~\rm{G}}\right)^{-4} ~,
\end{eqnarray}
where $P_b$ is in years (also see Fig.~\ref{fig:map}). Because any induced $f_2$ for sources considered in this section will be unmodeled in the timing model, the signal due to the orbital frequency of the binary will be reduced by a factor characterized by the transmission function \citep{bla84,mad13,haz19}. The transmission function quantifying the power absorbed due to spin-down, modeled as a quadratic function of time, goes as 
\be\label{eq:tf}
    \mathcal{T}_f \sim \left(\frac{T_{\rm{obs}}}{1.086~ P_b}\right)^6~,
\ee
where $T_{\rm{obs}} < P_b$ for {this equation, which is unitless, to hold} \citep{jen20}. By combining Eqns. \ref{eq:fn} and \ref{eq:tf} {where $\sigma_{\rm{rms}} = \mathcal{T}_f \times \sigma_{f_2}$}, we calculate the maximum $f_2$ that could evade detection over the given observation span {$T_{\rm{yr}}$ in years} using published timing residuals {$\sigma_{\rm{rms}}$ in seconds} and set a lower limit on orbital period 
\be\label{eq:lowerlimit}
    P_b \gtrsim 2.7~  \left(\frac{\tilde{M}}{10~M_{\odot}}\right)^{0.04} \sigma_{\rm{rms}}^{-0.12} ~ T_{\rm{yr}}^{1.08} ~.
\ee

\begin{figure}
    \epsscale{1.2}
    \plotone{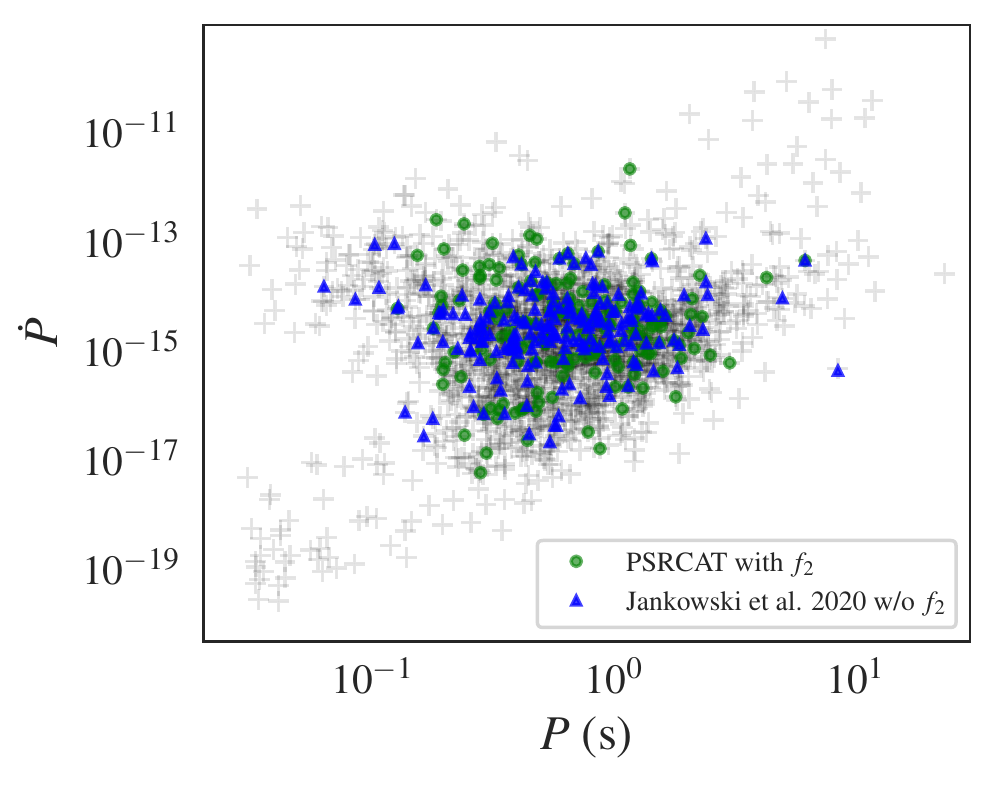}
    \caption{$P$-$\dot{P}$ diagram of identified candidates. The light gray crosses show all the known non-millisecond pulsars ($P > 30$\,ms) with published $f_1$ in \atnf. Unassociated isolated pulsars with published $f_2$ constraints are represented with green circles, and blue squares show those observed by \cite{jan18} without a published constraint on $f_2$. 
    }
    \label{fig:ppdot}
\end{figure}

\subsubsection{Pulsars with published constraints on $f_2$}

Next, we look at pulsars with published constraints on $f_2$. Again, we use the magnetic field to set an upper limit on possible orbital periods. We evaluate Eqn.~\ref{eq:fn} for the published $f_2$ to calculate the minimum orbital period that could be hidden for a given pulsar
{\begin{equation}
    \left(\frac{P_b}{100\,\rm{yr}}\right) \gtrsim \left(\frac{\tilde{M}}{10~M_{\odot}}\right)^{1/7} \left(\frac{f_0}{1\,\rm{Hz}}\right)^{3/7} \left(\frac{f_2}{10^{-22}\,\rm{s}^{-2}}\right)^{-3/7}~. 
\end{equation}}
Periods shorter than this limit will induce a higher $f_2$ than the published constraint, and are therefore unphysical for that pulsar.

\subsubsection{Candidates with published $f_3$ and/or $f_4$}

\begin{deluxetable}{lccc}
\tabletypesize{}
\tablecaption{Unassociated pulsars with published 
$f_3$ and/or $f_4$}
\tablewidth{0pt}
\tablehead{& \colhead{J1910+0517$^ a$} & J1913+1330$^ b$ & 
J1929+1357$^ a$
}
\startdata
$f_2$ (s$^{-2}$) & \phantom{$-$}7.5(6)$\times$10$^{-24}$ & 6(5)$\times$10$^{-27}$ & \phantom{.}1.00(2)$\times$10$^{-23}$ \\
$f_3$ (s$^{-3}$) & $-$1.8(3)$\times$10$^{-31}$ & $-$6.9(3)$\times$10$^{-33}$\phantom{$--$} & $-$4.2(4)$\times$10$^{-31}$ \\
$f_4$ (s$^{-5}$) & $-$2.6(3)$\times$10$^{-38}$ & \phantom{}7(1)$\times$10$^{-41}$ & \ldots \\
$\omega$ (rad) & \ldots & 0.11(6) & $<$0.52(3) \\
$P_b$ (yr) & \ldots & 0.5(3) & $>$3.0(6) 
\enddata
\tablecomments{Limits on orbital period and orientation based on published constraints on higher order frequency derivatives for unassociated pulsars in PSRCAT. Signs on frequency derivatives due to a binary would need to alternate positive and negative or all be the same sign to be physical; therefore, an estimate on orbital parameters could not be made for PSR~J1910+0517. \\
\footnotesize{a -- \cite{lyn17}}; 
\footnotesize{b -- \cite{bha18}}}  \label{tbl-higherorders}
\end{deluxetable}

With the availability of higher order frequency derivatives constraints, even more stringent limits on orbital parameters can be made. Comparing two frequency derivatives $f_n$ and $f_{n+1}$ gives a range of possible orbital periods and longitude of periastron
\be
    P_{b} = \left(\frac{-2\pi}{n+1}\right) \left(\frac{f_n}{f_{n+1}}\right) 
    \cot \left(\omega + \frac{n \pi}{2}\right).
\ee
While two frequency derivatives can be used to constrain a possible range of values for $P_b$ and $\omega$, using three derivatives constrains this range of values to a single estimate for each. There are three unassociated pulsars in \atnf ~with higher order frequency derivative constraints: PSR~J1929+1357 has a published $f_3$ constraint \citep{lyn17}, whereas PSRs J1910+0517 and J1913+1330 have published constraints on both $f_3$ and $f_4$ \citep{lyn17,bha18}. The more stringent limits on a possible binary in these systems are listed in Table~\ref{tbl-higherorders}.

\begin{figure}
    \epsscale{1.2}
    \plotone{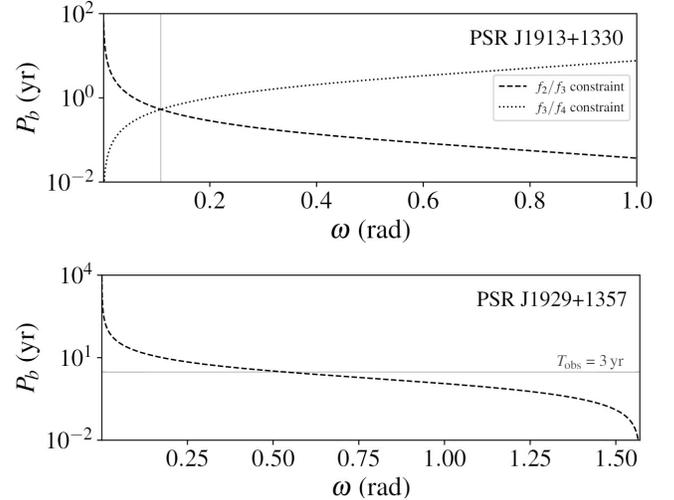}
    \caption{Limits on $P_b$ as a function of longitude of periastron for PSRs 1913+1330 and 1929+1357. Using constraints on higher order derivatives, we identify the range of combinations in longitude of periastron and orbital period that satisfy these constraints. In the top panel, J1913+1330 has published constraints up to $f_4$, which allows for more constrained orbital parameters compared to J1929+1357 in the bottom panel, which only has published derivatives up to $f_3$.
    The gray line in the bottom panel shows the timing baseline from \cite{jan18} {for comparison as a lower limit on a potential binary period}. {Note that using the ratio of frequency derivatives cancels out the companion mass, causing the limits calculated here to be independent of mass. }}\label{fig-1929}
\end{figure}

Depending on the longitude of the binary with respect to the periastron, all successive frequency derivatives due to circular binary motion should have either like or alternating signs, as can be seen in Eqn.~\ref{eq:fn}; this is not true for J1910+0517. Therefore it is unlikely that these two measurements are due to an unmodeled binary. However, this binary criterion is met for J1913+1330 and a more constrained orbital period can be determined. By comparing $f_2$, $f_3$, and $f_4$, we solve for a value of $\omega$ that gives a $P_b$ that satisfies all three measurements. We find a possible orbital period of $P_b = 0.5 \pm 0.3$\,yr with a longitude of periastron of $\omega = 0.11 \pm 0.06$\,rad. Without a constrained $f_4$ for J1929+1357, we cannot solve for both $P_b$ and $\omega$, but can identify combinations of the period and longitude of periastron that agree with the published constraints on $f_2$ and $f_3$. A plot of these two limits can be seen in Fig.~\ref{fig-1929}.

It is worth noting that these three pulsars show time-variable emission and are categorized as rotating radio transients (RRATs). While constraints on binary parameters can be obtained using measured frequency derivatives, confirming companions may be difficult given the transient nature of these pulsars {and the irregular timing of the pulses, although in some cases deeper observations may be able to reveal more regular emission \citep[e.g.,][]{mickaliger18}}.

\subsection{Constraining binaries within the population}

\begin{figure}
    \epsscale{1.2}
    \plotone{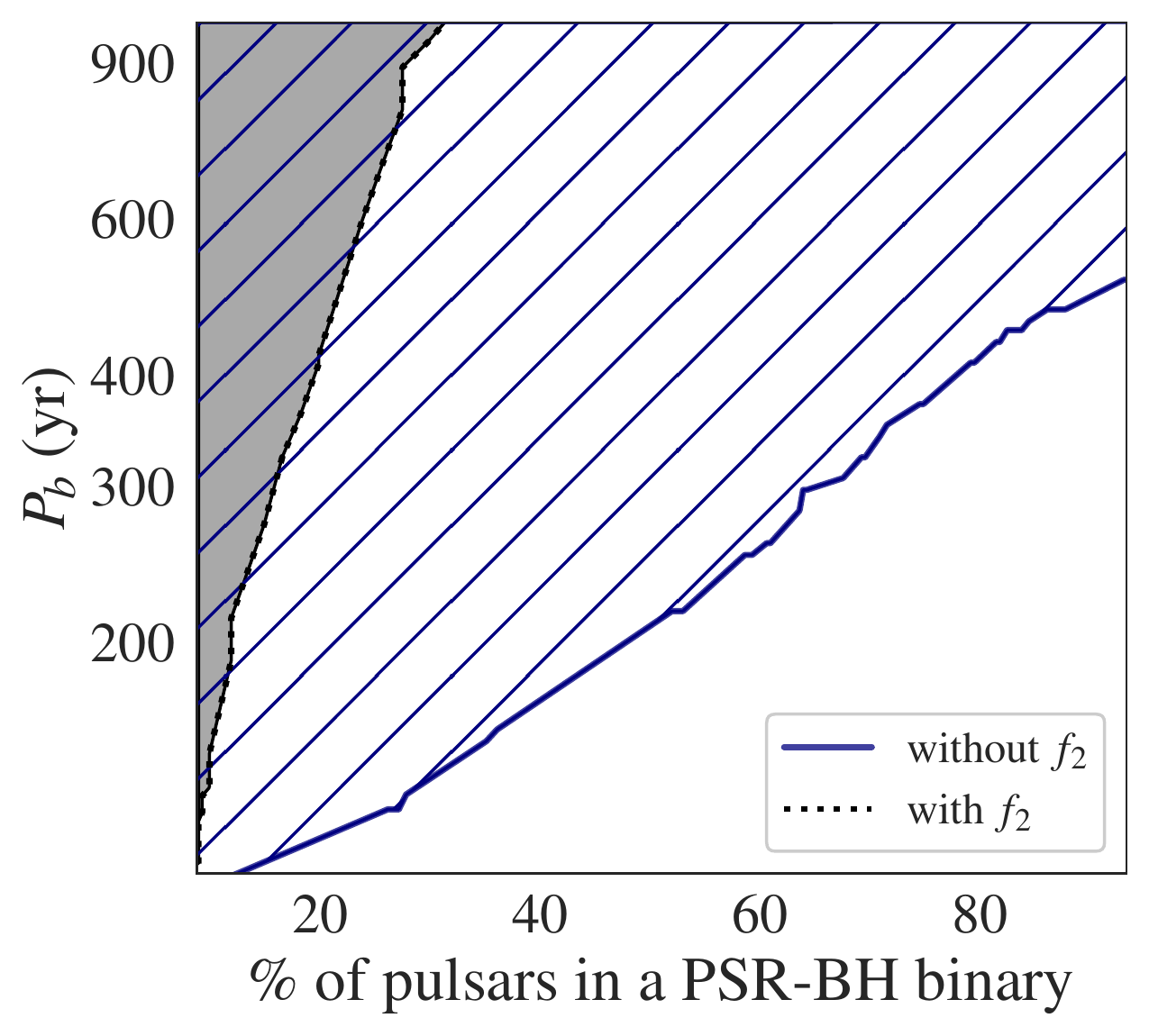}
    \caption{
    The percentage of isolated pulsars that could hide a pulsar-BH binary at a given orbital period, assuming the binary is edge-on with $M_{\rm{BH}} = 10$\,\msun. The shaded and hatched regions show where the likelihood level is within 95\%. The gray region includes pulsars with a published $f_2$, while the blue hatched region shows pulsars without published higher order frequency derivatives. As the percentage of pulsars with potential hidden binaries increases, the likelihood for a fixed period decreases. With the availability of higher order frequency constraints, the probable region decreases substantially. For instance, for pulsars with a published $f_2$ and $P_b=300\,$yr, we can exclude that more than 10\% of pulsars hide such a binary, while up to $\sim$60\% of pulsars without $f_2$ could hide such a binary.  
    }
    \label{fig-poisson}
\end{figure}

For pulsars with a published constraint on $f_2$, we apply our $f_2$ based limits to the full isolated pulsar population in \atnf. Due to the large number of isolated pulsars without a measured $f_2$ in \atnf, we focus on a subset of the population observed in \cite{jan18} without published $f_2$ measurements to ensure more uniform observations for the evaluation of the rms timing residuals and therefore calculate a lower limit on minimum orbital period using Eqn.~\ref{eq:lowerlimit}. To reduce the number of degrees of freedom, we assume the mean value of $2/\pi$ for {$\sin(\omega + n \pi/2)$} for this analysis.

Comparing Eqn.~\ref{eq:fn} to the published frequency derivatives in both populations above, we predict the fraction of sources that could hide a binary with some minimum orbital period based on published constraints on the frequency derivative. Assuming a Poisson distribution for the anticipated percentage of pulsars hiding a binary, we calculate the probable ranges for the population fraction hiding binary companions with a given minimum orbital period. Ranges are given for pulsars both with and without a published $f_2$, seen in Fig.~\ref{fig-poisson}. Note that these ranges are empirically derived from population parameters and are not physically motivated from binary formation.

By examining the percentage of pulsars with measured $f_2$ that could hide a binary with various orbital periods, we fit the following relation to our populations to yield an upper limit on the number of sources with orbital periods above a given minimum orbital period $P_{b;\rm{min}}$: 
\be\label{percent}
    \log_{10}{P_{b;\rm{min}}}  = \alpha + \beta x ~,
\ee
where $x$ is the percentage of the population that could hide  a corresponding $P_{b;\rm{min}}$ (in years). 
Applying a least-squares fit to the 177 PSRCAT sources with a published $f_2$ finds $\alpha = 2.17(5)$, and $\beta = 3.2(2)$. Similarly, for the 158 sources without $f_2$ this fit yields $\alpha = 1.95(2)$ and $\beta = 0.99(5)$; {with these values for $\alpha$ and $\beta$, we find} $\sim 35$\% as an upper limit in our sample that could hide a binary with $P_{b;\rm{min}} \geq 200$\,yr. Note that our analysis found 100\% of pulsars in \cite{jan18} could hide a binary with a minimum orbital period of $\sim$1\,kyr; therefore, Eqn.~\ref{percent} breaks down when investigating minimum periods above this value without a constraint on $f_2$.

\section{Discussion and Conclusions }

Building on historical work, we developed relations between effects on measured parameters and unmodeled binary motion. Using published constraints on higher order frequency derivatives, we identified a range of orbital periods and longitudes of periastron for one pulsar, and placed a more stringent constraint on these parameters for another pulsar. A third pulsar had published frequency derivatives that are not congruent with binary motion. 

Using limits on orbital period, we found that a long period binary could not be ruled out for roughly 30\% of pulsars with nonzero $f_2$ constraints, and that binary periods greater than $\sim1$\,kyr could not be ruled out for pulsars analyzed here without a constrained $f_2$. Sources that do not have published higher order frequency derivatives may still produce measurable constraints on these derivatives with further timing, particular through high cadence programs like CHIME and the future DSA-2000 \citep{chime, dsa2000}. The lower limit on orbital period given an observing time $T_{\rm{yr}}$ and a detected $f_2$ due to a binary can be determined from: 
\be
    P_b \geq 5.89 \times 10^{3} |f_{2; \rm{b}}|^{1/6} \left(\frac{T_{\rm{yr}}}{\sigma_{\rm{toa}}}\right)^{3/2} ~,
\ee
\\
\noindent where $\sigma_{\rm{toa}} \propto \sigma_{\rm{rms}} S_{\nu}^{-1} \sqrt{W (P - W)}/P$, $S_{\nu}$ is the frequency-dependent flux, and $W$ and $P$ are the pulse width and period respectively \citep{lor12}. For example, a 10\,$M_{\odot}$ binary companion in a 100\,yr orbit with a pulsar having a 10\,Hz spin frequency{, 10\,ms pulse width, and 100\,mJy flux at 843\,MHz} would be detectable with monthly 30\,min observations in $\sim$5\,yr assuming an instrument like UTMOST \citealt{low20}. With this observing strategy, higher order derivatives like $f_3$ and $f_4$ would be detectable in $\sim$7\,yr and $\sim$10\,yr respectively. 

{However, we note that this analysis assumes that timing is performed with sufficient cadence to capture any glitches, and that any timing variations due to changes in pulse shape or interstellar dispersion are small.  For the most part, the latter effects are only detectable for precision timing of millisecond pulsars \citep[e.g.,][]{jones17} used for detecting gravitational waves, and are unlikely to be significant for younger, noisier pulsars considered here.  Unmodeled glitches may set a floor for the detectability of any binary  to the extent that they are not already included in the empirical timing noise models used above.  In that case our analysis may be overly conservative, and so the true fraction of binaries that we cannot exclude could be lower than 30\%.  }

Follow-up of candidates with \textit{Gaia}, while excellent for shorter binary periods \citep{min18}, will heavily depend on binary orbital period and distance to the pulsar. For example, \cite{and22} utilize Data Release 3 astrometry to identify potential neutron star-BH binaries, with orbital periods of $\sim$0.9 -- 3.8\,yr. \textit{Gaia} lists an upper limit of 40\,yr on detectable orbits and therefore is not feasible for the orbital period ranges identified here, but could still be useful in finding stellar companions by identifying common proper motion pairs. 

Unmodeled binary motion will cause a deviation from linear proper motion. Multiple observations over time with instruments like the VLBA and ngVLA could hypothetically track the proper motion over time; however, the motions described here will induce deviations in proper motion that are smaller than published measurement errors. For example, a pulsar with a proper motion of 10\,mas/yr and a distance of 1\,kpc in a 20\,yr binary would see a change in proper motion of $\sim 10^{-5}$\,mas/yr after 5\,yr of observing. For this method to successfully follow up on candidates and complement timing constraints on proper motion in the line of sight, constraints on motion in the plane of the sky would require higher precision {of $\sim 10^{-3} - 10^{-2}$\,mas/yr}.

\acknowledgments
The authors would like to thank Nihan Pol and Joseph Lazio for useful discussions and manuscript comments. The authors are members of the NANOGrav NSF Physics Frontiers Center (awards \#1430284 and 2020265). 

\bibliography{psrbh}{}
\bibliographystyle{aasjournal}

\end{document}